# SFQ Bias for SFQ Digital Circuits

Vasili K. Semenov, Evan B. Golden, and Sergey K. Tolpygo, *Senior Member, IEEE*

*Abstract*—Superconductor electronics fabrication technology developed at MIT Lincoln Laboratory enables the development of VLSI digital circuits with millions of Josephson junctions per square centimeter. However, conventional DC and multi-phase AC biasing techniques already encounter serious challenges for scaling circuits above several hundred thousand junctions. In this work, we propose a novel AC-based biasing scheme for RSFQ-type logic families requiring DC bias. The major step toward this scheme is a superconducting AC/DC rectifier which we introduced at ASC 2014. We proposed to connect the rectifiers to "payload cells" via superconducting inductors with large inductance in order to reduce parasitic effects of flux quantization. Recently, we discovered that this powering scheme works even better at a much lower value of the inductance, when it is just sufficient to hold only one or two flux quanta in the inductive loop between the converter and the payload. In this case, flux quantization in the loop becomes beneficial because the value of current fed into the payload is defined by the value of the coupling inductance. Therefore, our AC/SFQ converter powers the payload cell by single flux quanta rather than by DC current. Such mode of operation is extremely energy efficient because energy is used only to recover the flux quantum consumed by the cell during the logic operation. We present designs of AC/SFQ converters comprising an AC/DC rectifier and a current conditioning circuit which we termed an SFQ filter. We also present test results and demonstrate AC/SFQ powering a payload circuit using circuits fabricated in a new, 150-nm node of Lincoln Laboratory fabrication technology using self-shunted Nb/AlO$_x$-Al/Nb Josephson junctions with 600 μA/μm$^2$ critical current density and 200 nm minimum linewidth of inductors.

## I. Introduction

A noticeable discrepancy between anticipated and achieved results of recent projects in the area of superconductor digital electronics indicates existence of hidden impediments. One of them is likely to be related to limitations of the existing biasing schemes. RSFQ technology [1] and its recent eRSFQ and ERSFQ flavors [2], [3] are based on the conventional DC biasing technique. The latter serves well at modest integration levels but it could hardly be scaled up above one million Josephson junctions per chip level of integration because of high total bias currents which might exceed hundreds of amperes. Current recycling [4] - [6] allows for a reduction of bias current by one order of magnitude. AC biasing techniques do not require large bias currents. However, in RQL [7], AQFP [8] and some other logic types, AC bias current also serves as a global clock signal. Such rigid timing is highly impractical and leads to significant hardware overhead. In some cases, up to 90% of gates operate merely as elements of the shift registers transferring the processed data through the processor [9]. The main goal of this work is to develop a new scalable AC biasing scheme where a single-phase AC current powers circuits without serving as a mandatory synchronization signal. In this paper we show how to use AC power to feed DC-powered cells.

In [10], we proposed that a magnetically biased SQUID can be used as a diode because of a significant difference between its positive and negative critical currents; see Fig. 1 in [10]. Two SQUIDs with opposite magnetic biases, as shown in Fig. 2 in [10], can serve as an AC/DC converter. We also proposed to stabilize the DC current generated by the AC/DC converter by a current-limiting Josephson junction and to depress the undesirable quantization effect by means of a wire with a very large inductance connected in series with the limiting junction; see Fig. 4 in [10]. Recent investigations have shown that our powering scheme works much better without the limiting junction and at a much lower value of the inductance connecting AC/DC converter with an SFQ payload. In this case, the payload circuit is fed with single flux quanta rather that with a DC current. Circuit design and experimentation details are presented below.

## II. Design

### A. Basic Concept of Superconductor AC-DC Converter

The new biasing concept can be explained in two steps. The first step is illustrated by a circuit in Fig. 1a, composed of an AC/DC converter, relatively small inductor, and an ideal phase source, PD, that emulates an SFQ payload circuit. The schematics of the AC/DC converter is shown inside a gray box in Fig. 1. It comprises two identical asymmetric SQUIDs with different positive and negative critical currents, which are connected in "opposite" directions for the AC current. This asymmetry is created by applying a constant magnetic flux bias $\Phi_0/4$ to one SQUID and $-\Phi_0/4$ to another SQUID, corresponding to $\pi/2$ and $-\pi/2$ phase shifts, respectively, using oppositely winded trans-formers formed by inductors L1A and L2A and a current-carrying wire, as shown in Fig. 1. Both SQUIDs are in the superconducting state until the induced AC current value is below the lower of two critical currents. At higher current values one of the SQUIDs becomes "resistive" and the induced current should bypass it via the superconducting load. In reality, the passage of a single flux quantum through the resistive SQUID turns it back into superconducting state, but the induced current continues to flow into the load. The same process repeats in the second SQUID after the AC current reverses its direction. As a result, at each half-period of the AC bias, a fraction of the current is squeezed into the load in the same direction. The remaining AC current in the converter is insufficient to turn either SQUID into the resistive state.

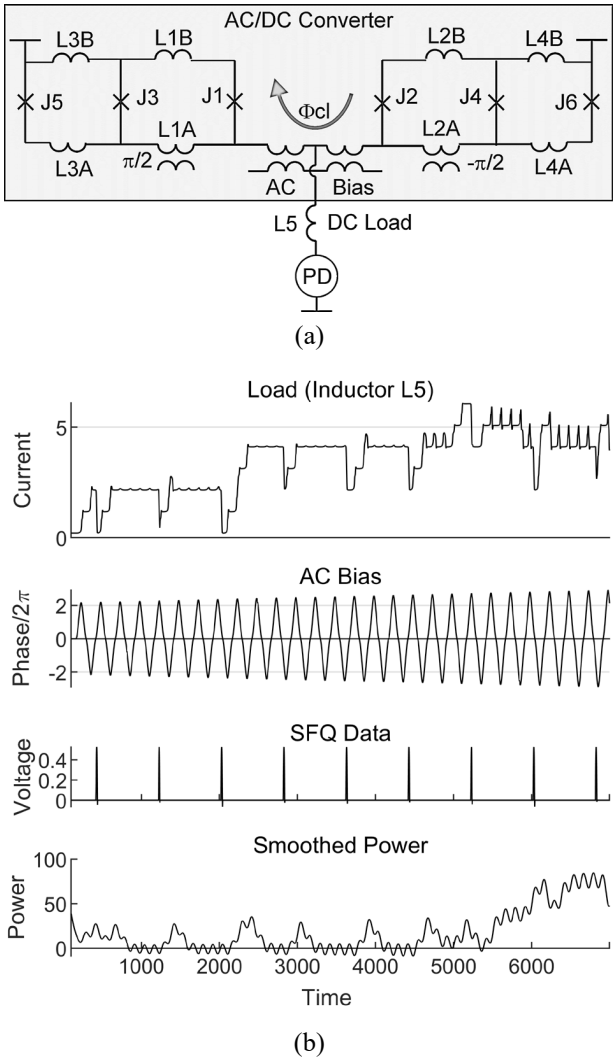

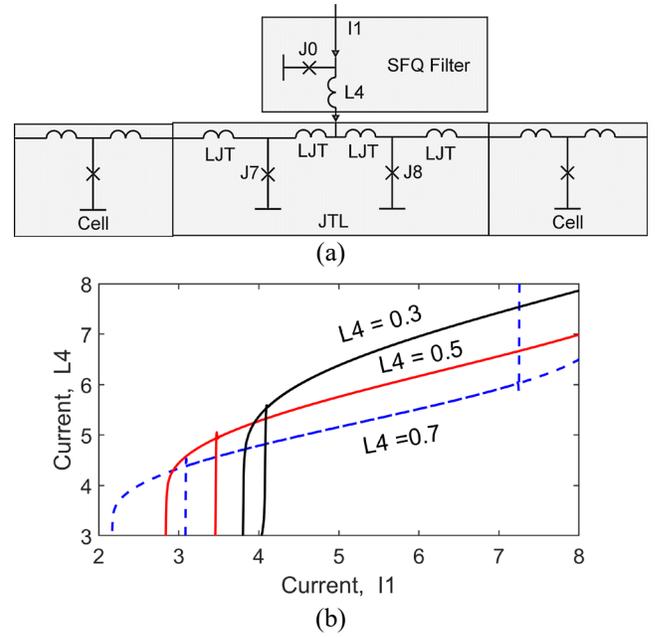

Fig. 1. Schematics (a) and operation (b) of AC/DC converter as a current source at the following set of dimensionless, PSCAN, parameters: Ic1=Ic2=5, Ic3=Ic4=Ic5=Ic6=3.75, L1A=L2A=0.25, L1B=L2B=0.06, L3A= L4A=0.168, L3B= L4B=0.07, L5=2.32. All circuit optimizations in this work were done using normalized, dimensionless PSCAN units [11]. To convert to the regular units, the following set of PSCAN units could be used: $I_U$=0.125 mA for currents, 1.7 mV for voltages, $L_U$=2.64 pH for inductances, and flux quantum $\Phi_0$ for magnetic flux and phase. The unit of power is $P_U$=$I_U \cdot \Phi_0$ per one period of AC bias.

Fig. 2. Schematics (a) and operation (b) of an SFQ filter loaded with a JTL. Calculations at J0=2 have been performed at a temporarily increased critical currents of junctions J7 and J8, which are nominally equal to 2 current units.

Numerically simulated traces shown in Fig. 1b illustrate the described dynamics of the circuit components at a slowly growing amplitude of the applied AC bias. The simulation shows that the load current via inductor L5 (the upper trace) is always positive and the converter works as a rectifier of AC current. The value of the load current is quantized and is not affected by small changes of the AC bias. The load current remains within a range between 3 and 5 PSCAN current units ($I_U$), or between 375 to 625 μA, even if the amplitude of AC bias changes by about 20%. However, the current jumps between the quantization levels if the AC bias amplitude changes too much. The jumps between quantized current levels in L5 are also initiated by the release of SFQ pulses into the payload circuit simulated by ideal phase source PD, as shown by the third trace in Fig. 1b. These short-lived disturbances visualize the about one period of AC bias refractory time when the circuit is not ready for the next SFQ pulse. In other words, the refractory time sets the minimum allowed time interval between adjacent SFQ pulses.

At the second step, we demonstrate how to condition or stabilize the load current flowing via inductor L5. Fig. 2a illustrates that a single-junction SQUID comprised of junction J0 and inductor L4 performs this job and works as a filter. For example, at L4 = 0.5$L_U$, this filter squeezes the applied current, I1 from the range from 3.5 to 8$I_U$ into the range from 5 to 7$I_U$ in L4. We suggest to call the circuit an SFQ filter.

We note, that the maximum bias current flowing into the load JTL in Fig. 2a may exceed the sum of critical currents of its two Josephson junctions in order to feed "bias-less" cells connected to the edges of the JTL. In this case, the applied current spreads out beyond the JTL and feeds two cells which do not have their own bias sources. In Fig. 2a, the cells are emulated by single Josephson junctions, but they could be real RSFQ-like "bias-less" gates in a more practical circuitry. The optimal bias current of the composite payload, a JTL and bias-less cells, is matched to the output of the SFQ filter. Note that the effective inductance of the SFQ filter is composed of inductance L4, inductances of JTL inductors, and inductances of junctions J7 and J8.

### B. AC/SFQ Converter

A proposed SFQ biasing scheme of SFQ circuits is illustrated in Fig. 3a that shows a fragment of an SFQ circuitry which includes AC/SFQ converters, composed of an AC/DC converter and SFQ filter, loaded by JTLs and a couple of cells. All the essential components have been discussed above. Fig. 3b proves that their strong interactions do not ruin the whole concept. The upper trace in Fig. 3b shows the AC bias applied to the converter and simulated by an AC phase source. The amplitude of the source is constant, but one period of the AC bias is intentionally missing. The second and third, from the top, panels show voltage pulses and phase drops corresponding to the propagation of three fluxons along the loading JTL. Besides SFQ pulses generated by junctions J7 and J8, we show pulses generated by junction JX located in the preceding Cell located on the left side of the JTL in Fig. 3a.

Initially all currents are set equal to zero. The first period of AC bias current "charges" the JTL, while the second period is absent to demonstrate the principle of operation. Despite the lack of power, the first fluxon rapidly passes through all three junctions, resulting in an almost full coincidence of three leftmost pulses. Such behavior of our SFQ biasing circuit is similar to an Uninterruptable Power Supply (UPS) that keeps current sufficient to transfer one fluxon through the circuit even in the absence of external power. The passage of the fluxon discharges the JTL biasing inductor L4 and, because of the power interruption, caused by the absence of the second AC period, remains discharged. Earlier we banned the entrance of the new fluxon before the power is restored. Now we violate the rule and send the second fluxon in the discharged JTL. The fluxon passes via JX but stops before J7, and continues its motion only after the next period of the AC power restores current in L4 and the JTL bias.

The described functionality is similar to the behavior of a coincidence junction or Muller C-element discussed, for example, in [1], see Fig. 24 in [1]. Indeed, the C-element signals to the output terminal the arrival of the second of the two input signals. In our case, one of the two logic input signals is substituted by the waves of the AC bias. In the normal operation, AC waves precede SFQ data pulses and do not affect their propagation. However, if a data pulse arrives first, then it waits for the next wave of the AC bias. Here we demonstrated that the punishment for the premature arrival is not too high. Additionally, we show that, if desired, the AC bias current could be involved in timing the SFQ circuitry.

The third fluxon in Fig. 3 rapidly crosses the now re-charged JTL due to normal operation of the AC/SFQ bias source. The fourth panel from top shows that the output currents are quantized and that the current via L4 is indeed better stabilized than the current via inductor L5. The fifth panel shows phase drops on two junctions, J1 and J2, of the AC/DC converter and highlights that $2\pi$ phase leaps on the converter junctions take place only during active charging and recharging of the JTL bias.

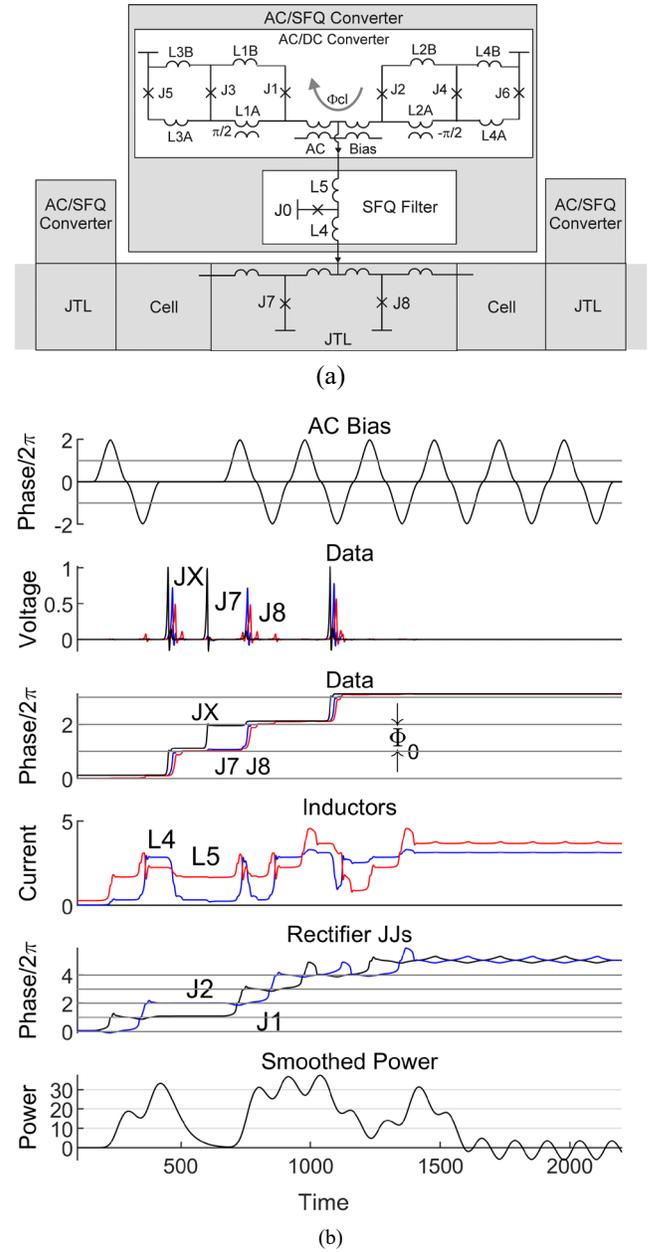

(a)

(b)

Fig. 3. Fragment of SFQ circuitry with three AC/SFQ converters, three JTLs and two cells (a) and its operation (b). For the Voltage Data panel, black curve shows JX pulses, blue color shows voltage on junction J7, and red curve shows voltage on junctions J8. For the Phase Data panel, black curve shows JX pulses, blue color shows voltage on junction J7, and red curve shows voltage on junctions J8. For Inductors, blue curve is current in L4 and red curve is current in L5. For Rectifier JJs, the y-axis grid intervals correspond to $2\pi$ or $\Phi_0$, black curve is phase drop on J2 and blue curve is phase drop on J1.

The recharge is required after each SFQ data pulse, because passing a data pulse is equivalent to a withdrawal of the flux quantum from the loop with inductor L4. The recharge is activated when current via L4 drops to almost zero, and the entire DC current from the lower part of the converter begins to flow via J0. This continues until the current in J0 exceeds its critical value and the new flux quantum enters the loop with inductor L4 via J0. At this point, the recharge is completed and the circuit is ready for the next SFQ pulse.

The bottom, sixth, panel shows the smoothed power consumed from the AC bias. The power is calculated as a sum of two products of currents and voltage drops on ideal phase sources emulating the applied AC bias. However, the power itself is difficult for a visual perception and we passed the calculated power via a built-in running averaging window [11]. The plot is normalized by power dissipated by vortices passing a Josephson junction biased by one $I_U$ current at one vortex per clock period rate. In such units, the numerical value of power would correspond to total bias current that passed by a vortex per clock period.

The right side of the plot corresponds to the idle mode, when the AC/SFQ converter operates without two-pi jumps and it is easy to see that the average power is close to zero. At the left side of the plot, peaks crudely corresponding to the power of a vortex passing via current equals to sum of critical currents of all junctions in the AC/SFQ converter. The traces in Fig. 3b illustrate a strong interaction between circuit components operating in a single-flux-mode. This mode is rather difficult for analytical calculations but quite easy for the numerical simulation and optimization using PSCAN2 software package [11].

*C. Discussion of More Advanced Implementations*

It may appear that the new biasing technique is a bit complicated and causes a significant junction and area overhead. Fortunately, this is only partially true because, even for applications to small-scale circuits which do not require large bias currents, it allows, *e.g.*, to reduce the total number of required bias wires. Most importantly, this technique is the only viable option to bias circuits with about or over ten million junctions. Secondly, we recommend implementing the new bias technique with a prospective flavor of bias-less logic cells operating without any explicit dc bias current sources. A few examples of such cells optimized for the use with pi-shifters are shown in Fig. 4. Such cells are rarely discussed because ideally they would require implementing, still-exotic, pi-shifters based on ferromagnetic JJs; see, *e.g.*, [12], [13]. Fortunately, solid-state pi-shifters can be substituted by magnetic coupling of cells to an appropriate DC current. Such magnetic phase shifting has been known since at least 1982 [14]. In this case, all required pi-shifters can be implemented as transformers fed by a single current source.

Our numerical simulations showed that pi-biased cells require lower bias currents. This is because the pi-shift converts the core part of the cells into a symmetrical two-junction SQUID with the maximally depressed critical current; see, *e.g.*, Fig. 4a. As noticed above, small bias currents can be received via adjacent JTLs connected with the inputs and the outputs of the cells. This technique is known and was demonstrated, for example, in Fig. 4 in [15]. As an example, Fig. 4b shows a schematics of pi-biased T flip-flop which lacks two junctions in comparison with the standard implementation. The functionality of these two junctions will be executed by JTLs connected to outputs FN and FP, as discussed above.

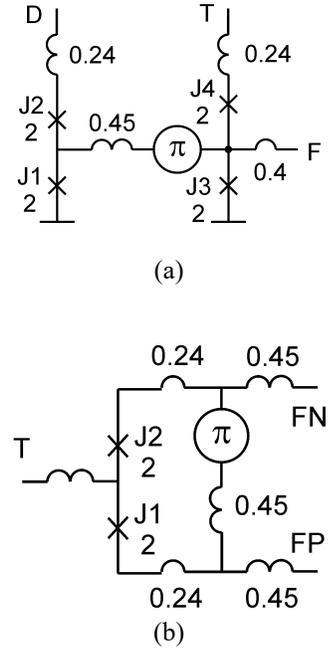

Fig. 4. D flip-flop (a) and T flip-flop (b) with pi-shifters. All circuit parameters are given in dimensionless PSCAN units.

More biasless cells and cells with reduced values of bias currents need to be invented to reveal all benefits of the SFQ biasing technique. Such cells will require AC/SFQ converters with different load currents. The required load currents can be readily obtained by properly scaling the AC/SFQ converter shown in Fig. 1, which for this purpose was optimized using normalized, dimensionless units.

The most convenient implementation of the new bias technique is in advanced fabrication technologies with two vertically stacked layers of Josephson junctions [13], [16], [17], allowing vertical integration of AC power converters and logic circuitry, thus minimizing area overhead. We expect the proposed biasing technique to flourish when these technologies become widely available.

III. LAYOUT AND FABRICATION

The AC/SFQ biasing project was initiated as the development of demonstration circuits for a new 150-nm node of the SC1 fabrication process developed MIT Lincoln Laboratory [18] and presented at the ASC 2020 [19]. The process uses self-shunted Josephson junctions with critical current density of 600 µA/µm$^2$ [20] and allows for circuit densities over $10^7$ JJs per cm$^2$. In all designs, resistors were used only for circuit biasing. Extraction of self and mutual inductances was a new and challenging exercise. Fortunately, Inductex software package [21], [22] was able to handle the layouts even at deep submicrometer dimensions.

Fig. 5 shows schematics and layouts of the fabricated and successfully demonstrated AC/DC converter and load JTL. The schematic shown in Fig. 5a is similar to the one suggested in [10]. Please note that the test circuit in Fig. 5a does not have junction J0, i.e., the SFQ filter described in II.A was not implemented. The older schematics were used because the cells had been submitted for fabrication before the full optimization was completed. The cells occupy 7 μm x 12 μm area. Minimum linewidth of inductors in the cells is 0.2 μm. The red ellipses in Fig 5b highlight transformers that are the most challenging circuit components. Their miniaturization is limited by the required mutual inductances. About 5 mA of current should be applied at implemented dimensions to achieve the target magnetic biases. In order to reduce the lengths of the transformers by a factor of two, the value of the required current should be increased to 10 mA. A further decrease of dimensions requires new layout and technology solutions.

## IV. MEASUREMENTS

Our measurements confirmed the presented concept. Testing has been carried out in a low-frequency probe immersed in liquid helium, using an Octopux setup [23]. The measured circuit includes an AC/SFQ converter shown in Fig. 5. Conventional auxiliary DC/SFQ and SFQ/DC converters with separate DC biases were used to generate and detect SFQ pulses.

The testing procedure, see Fig. 6, examines the propagation of SFQ pulses through the circuit with and without AC bias. To simplify the measurement, the sine-wave AC current is substituted by rectangular pulses shown as upper traces in Figs. 6a and 6b. In the first "sampling," the AC bias current goes from zero to its maximal positive value. In the second event, the current returns to zero. In the third event, the current goes from zero to its maximal negative value. In the fourth event, the current returns to its initial zero value. So, the application of one period of AC bias requires four sampling events. The application of SFQ data pulse, see the middle traces labeled "Input Pulses" in Figs. 6a and 6b, is composed of only two events. In the first event, the input current rises from zero to its maximal positive value. In the second event, the current returns to its initial zero value. The output, the lower traces in Fig. 6, is monitored in each sampling event. The observation of each SFQ pulse is detected by the toggle of the output voltage. The AC and input pulses are generated in different sampling periods in order to distinguish between their impacts on the output.

A number of test patterns have been applied to the circuit at numerous values of all used biases. Fig. 6 shows the two most informative patterns. The upper pattern in Fig. 6a exhibits conventional behavior when all input pulses pass to the output because at least one period of AC current precedes the data pulses. This indicated that the AC/SFQ biasing scheme works as expected. The lower pattern, Fig. 6b, shows more exotic behavior when two input pulses are sequentially applied while the AC bias is off. The lower trace clearly shows that the first pulse passes to the outputs without any delay. However, the second pulse stays in the circuit and awaits the completion of one period of AC current.

We also investigated margins of the circuit operation with respect to the AC current amplitude, $I_{AC}$ and the common DC

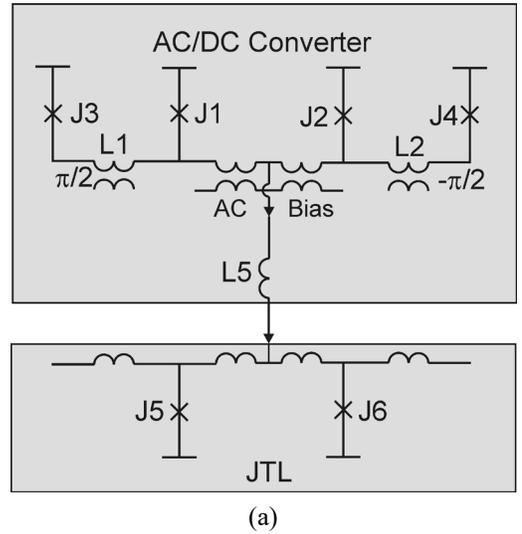

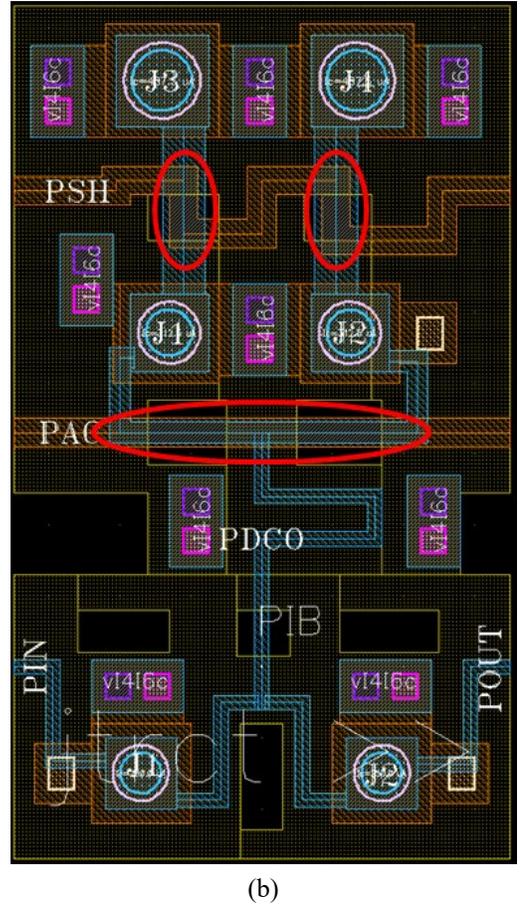

(b)

Fig. 5. Schematics (a) and layouts (b) of the fabricated AC/DC converter feeding a JTL. Essential parameters are $Ic1 = Ic2 = 0.31$ mA, $Ic3 = Ic4 = 0.49$ mA, $Ic5 = Ic6 = 0.25$ mA, $L1 = L2 = 1.24$ pH, the mutual inductance of the AC coupling transformer $M_{ac} = 2\times0.42$ pH, and the mutual inductance in the $\pm\pi/2$ flux bias transformers is $M1 = M2 = 0.1$ pH. The primaries of the AC power transformer and the magnetic flux bias transformers are marked PAC and PSH, respectively. All the transformers are marked by red ellipses.

current, $I_{PSH}$ in transformers creating equal and opposite phase shifts (flux bias) in the SQUIDs of the AC/SFQ converter. The typical margin plots are shown in Fig. 7 and Fig. 8. Fig. 7 shows the color-coded voltage response of the output SFQ/DC converter to the data SFQ pulses produced by the input DC/SFQ converter at time steps shown by the solid vertical lines, measured at different $I_{AC}$ values. Similar to Fig. 6, the output toggles between a low voltage level, coded red, and a high voltage level, coded blue, upon arrival of each data pulse. It can be seen that the margins of the correct operation are very wide, from about 7 mA to about 14 mA or about ±50%.

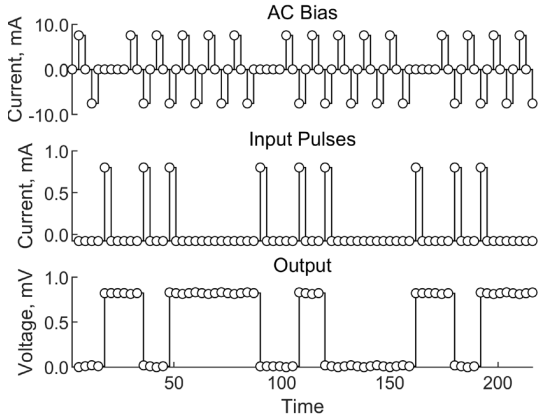

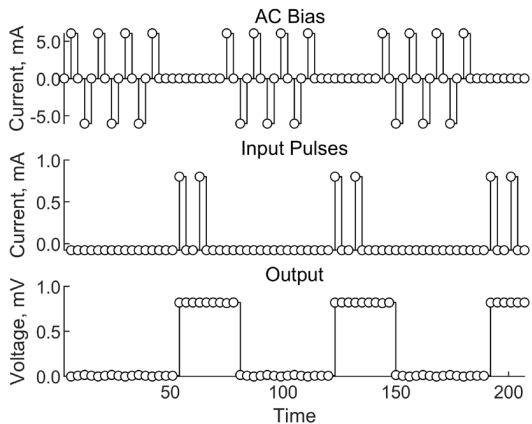

Fig. 6. Conventional (a) and more exotic (b) test patterns. Time is measured in the number of sampling events. Each sampling event is highlighted by a circle marker and lasts about 1 ms.

Fig. 8 shows the cumulative margins of the circuit operation with respect to varying both the $I_{AC}$ and the phase shifting (SQUIDs flux biasing) current $I_{PSH}$. The margins with respect to $I_{PSH}$ are ±20%. If we assume that the margins are centered at exactly $\pi/2$ ($\Phi_0/4$) phase shifts in the SQUIDs, the center value $I_{PSH}$ = 5.85 mA implies that the mutual inductances M1 and M2 are 0.088 pH or about 12% lower than the value simulated using InductEx. It should also be noted that only a part of the very wide $I_{AC}$ margins corresponds to the truly energy efficient mode of operation when each flux quantum consumed by the payload circuit is replenished by a single flux quantum taken from the AC/SFQ converter. At large AC amplitudes, the converter generates extra flux quanta which are not needed for the operation and get rejected.

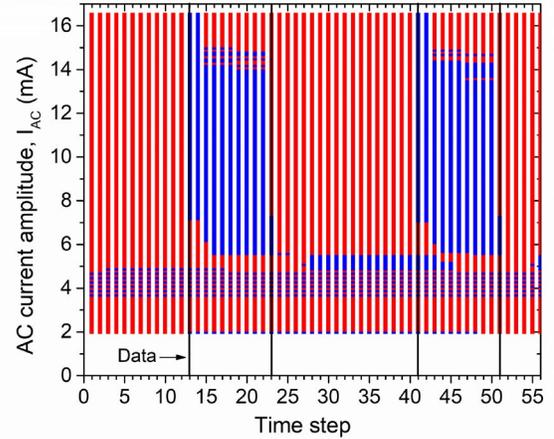

Fig. 7. The AC current amplitudes $I_{AC}$ applied to the circuit in Fig. 5 to verify its operation at a fixed value of the $I_{PSH}$ providing about ±$\Phi_0$/4 flux biases to the SQUIDs. The Data pattern shown by solid vertical lines corresponds to data pulses arriving at time steps 13, 23, 41, and 51. The $I_{AC}$ band corresponding to the correct operation, i.e., the proper toggling of the output voltage between low (red) and high (blue) levels upon arrival of each Data SFQ pulse, spans from about 7 mA to 14 mA. The $I_{AC}$ was incremented in 0.1 mA steps.

## V. CONCLUSION

We have suggested a new concept of powering of superconductor circuits with very large scales of integration. The concept is based on utilization of the novel AC/SFQ converters. The role of the converter is to supply the payload circuit with single flux quanta rather than current or voltage. The application of fluxons is partly similar to the conventional application of the DC bias current. The main difference is that biasing by SFQ results in the bias current temporally vanishing after each logic event executed by the payload. It is automatically restored in the converter after a short refractory time. Conventional AC/DC conversion in Josephson junctions is known since 1977 [24]. However, this effect is not practical for application as a power source.

The new biasing technique demonstrates significant advantages:

a) Due to the flux quantization, the value of bias current is accurately defined by the inductance of the superconductor inductive components connecting the payload with the converter. In the modern fabrication technologies, inductors are well reproducible components of superconductor circuits.

b) Parasitic ringing effects are strongly depressed when the payload bias current is close to zero, which is highly beneficial for increasing the allowed spread of circuitry parameters.

c) AC power currents do not have to provide global timing. The AC power frequency may differ from the clock frequency. The best energy efficiency is achieved if the AC power frequency exceeds the clock frequency. A moderate decrease of the energy efficiency could be traded in for a factor of two or three lower AC power frequency [10].

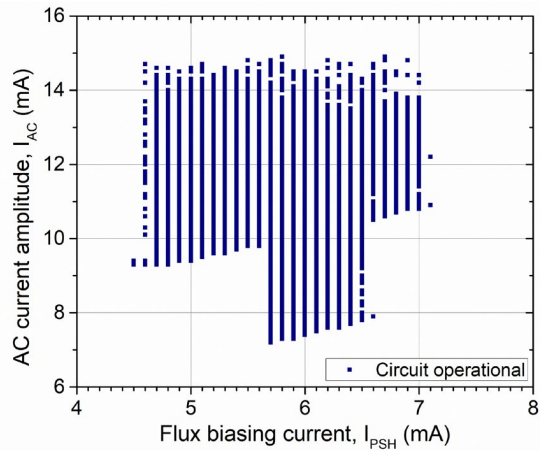

Fig. 8. Margins of operation of the AC/SFQ converter powering the JTL payload; see Fig. 5. For the margins search, both the AC current amplitude and the dc flux biasing current in PSH line were incremented in steps of 0.1 mA and various data patterns were applied to verify the circuit operation.

The suggested AC/SFQ powering concept has been experimentally verified using relatively simple payload circuitry. The experiments have demonstrated wide spread of allowed amplitudes of the AC bias and phase shifting currents. In the experimentally demonstrated case, the implemented AC/SFQ converter replenished one flux quantum consumed by the payload cell at each operation. Extension of this concept to SFQ biasing of more complex gates and gates with fan out is certainly possible and will be subject of future work.

It is expected that the proposed AC/SFQ converters will be implemented using the advanced multi-layer technology with two layers of JJs. A group of these layers, including one of the junction layers, will be reserved for on-chip power distribution – for the AC/SFQ converters and striplines delivering AC power, and for wires delivering pi-shifting magnetic biases.

Compact pi-shifters, *e.g.*, based on magnetic Josephson junctions, would be an essential addition to prospective VLSI superconductor circuits.

ACKNOWLEDGMENT

We are thankful to Sergey Rylov and Yuri Polyakov for fruitful discussions, to Alex Kirichenko and Pavel Shevchenko for their help with PSCAN2 software package, and to Conrad Fourie for his help with InductEx software package. The circuits were fabricated at MIT Lincoln Laboratory, using the 150-nm process node of the SC2 process. We are grateful to the MIT LL fabrication team, especially to Vladimir Bolkhovsky, Ravi Rastogi, and Scott Zarr for overseeing the wafer fabrication process. We also like to thank Leonard Johnson and Mark Gouker for their interest in and support of this work.

This research is based upon work supported by the Under Secretary of Defense for Research and Engineering under Air Force Contract No. FA8702-15-D-0001. Any opinions, findings, conclusions or recommendations expressed in this material are those of the author(s) and do not necessarily reflect the views of the Under Secretary of Defense for Research and Engineering.